\documentstyle[12pt,psfig]{article}  
 
\newcommand{\labell}[1]{\label{#1}}  
\newcommand{\reef}[1]{(\ref{#1})}

\def\IR{{\hbox{{\rm I}\kern-.2em\hbox{\rm R}}}} 
\def\IB{{\hbox{{\rm I}\kern-.2em\hbox{\rm B}}}} 
\def\IN{{\hbox{{\rm I}\kern-.2em\hbox{\rm N}}}} 
\def\IC{\,\,{\hbox{{\rm I}\kern-.59em\hbox{\bf C}}}} 
\def\IZ{{\hbox{{\rm Z}\kern-.4em\hbox{\rm Z}}}} 
\def\IP{{\hbox{{\rm I}\kern-.2em\hbox{\rm P}}}} 
\def\IH{{\hbox{{\rm I}\kern-.4em\hbox{\rm H}}}} 
\def\ID{{\hbox{{\rm I}\kern-.2em\hbox{\rm D}}}} 
\def\II{{\hbox{\rm I}\kern-.2em\hbox{\rm I}}}

\begin{document}  
  
\newpage  
\bigskip  
\hskip 4.7in\vbox{\baselineskip12pt  
\hbox{hep-th/0011166}}  
  
  
\bigskip  
\bigskip  
\bigskip  
\bigskip  
\bigskip  
\bigskip 
  
\centerline{{\Large \bf Probing Some ${\cal N}=1$ AdS/CFT RG Flows}}
  
\bigskip  
\bigskip  
\bigskip  
\bigskip 
  
\centerline{\bf Clifford V. Johnson, Kenneth J.
  Lovis, David C. Page}

\bigskip  
\bigskip  
\bigskip

\centerline{\it Centre  
for Particle Theory}
  \centerline{\it Department of Mathematical Sciences}  
\centerline{\it University of  
Durham}
\centerline{\it Durham, DH1 3LE, U.K.}  

\centerline{$\phantom{and}$}  

\bigskip

\centerline{\small \tt  
  c.v.johnson@durham.ac.uk, k.j.lovis@durham.ac.uk, d.c.page@durham.ac.uk}  
  
\bigskip  
\bigskip  
\bigskip  

  
\begin{abstract}  
  \vskip 4pt 
  
  We present the results of probing the ten dimensional type~IIB
  supergravity solution corresponding to a renormalisation group flow
  of supersymmetric $SU(N)$ Yang--Mills theory from pure gauge ${\cal
    N}{=}4$ to ${\cal N}{=}1$ with two massless adjoint flavours.  The
  endpoint of the flow is an infrared fixed point theory, and because
  of this simplicity of the theory, the effective Lagrangian for the
  probe is very well--behaved, having no zeros or singularities in the
  tension, and a smooth potential, all of which we exhibit.
  Specialising to the locus of points where the potential vanishes, we
  also characterise a part of the Coulomb branch of the ${\cal N}{=}1$
  theory. The simplicity of the gauge theory physics allows us to
  isolate and emphasise a key holographic feature of brane probe
  physics which has wider applications in the study of geometry/gauge
  theory duals.
\end{abstract}  
\newpage  
\baselineskip=18pt  
\setcounter{footnote}{0}  
  
  
\section{Introduction}  
One of the examples of the AdS/CFT correspondence\cite{malda,gkp,w1}
is a duality between ${\cal N}=4$ supersymmetric $SU(N)$ Yang--Mills
theory (at large $N$ and strong 't Hooft coupling) and type~IIB string
theory propagating on AdS$_5{\times}S^5$. In this strong form (and in a
number of weaker forms), is a remarkable relationship between the
dynamics of gauge theory and geometry. It points to the possibility of
developing many useful tools for application to the study of gauge
theories  relevant to very important physical phenomena, such as
confinement, the phenomenology of baryonic matter at high temperature
and density, {\it etc}.

One of the many avenues of investigation being explored in order to
make contact with such exciting physics is the issue of deforming the
correspondence by switching on relevant operators, so that it
represents the renormalisation group (RG) flow from ${\cal N}=4$
supersymmetric pure Yang--Mills theory in the ultraviolet (UV) to
${\cal N}=2,1$ or~0 supersymmetric Yang--Mills theories of various
sorts, in the infrared (IR). A number of supergravity solutions have
been constructed with such dual
interpretations\cite{gppz1}--\cite{kw}.
  
The supergravity solutions interpolate between AdS$_5\times S^5$ at
$r=+\infty$, the dual of the ${\cal N}=4$ theory (with strong 't Hooft
coupling in the UV; here $r$ is a suitably chosen radial coordinate of
AdS$_5$), and a solution in the interior at $r=-\infty$, dual to the
new gauge theory obtained after relevant perturbation and flowing to
the IR.

Recent work\cite{jpp} on other gauge/geometry situations have shown
that it is quite fruitful to focus on the physics which may be
obtained from studying the behaviour of a single probe D--brane (of a
suitable variety) in the supergravity background\footnote{For a
  collection of pedagogical studies with a focus on these techniques,
  see ref.\cite{primer}.}. This has led to the discovery of
significant modifications of the naive supergravity geometry, and
bridges the gap between the pure supergravity technology and that of
the full superstring theory which has yet to be developed for
propagation in these backgrounds\footnote{See also
  refs.\cite{cvj,cvjlaur,cvjagain, bertolini} for extensions and further study of
  the approach of ref.\cite{jpp}. Other recent appearances of the
  enhan\c con phenomenon may be found in refs.\cite{enhanced}.}.
Those investigations have uncovered a new rationale and mechanism for
removing troublesome singularities in geometries dual to gauge
theories\footnote{See ref.\cite{gub2} for a review of the
  singularities which occur in the RG flow context, and a proposal for
  their classification into physical and non-physical within the
  context of supergravity.}. The resulting ``enhan\c con'' geometry,
made of smeared branes, is consistent with the dual (or accompanying)
gauge theory, and often sheds new light on it.

The RG flow geometries are typically derived in the context of ${\cal
  N}=8$ supergravity in five dimensions, which is believed to be a
consistent truncation of of type~IIB supergravity. There are not many
complete ten dimensional ``lifts'' of these solutions known, and so
the probe techniques (which rely on having the full geometry, since
the probe brane couples to fully ten dimensional fields) have not been
applied to the study of the many RG flows which are known.

This situation is changing. Soon after the presentation in
ref.\cite{pw1} of the complete ten dimensional lift of flows to
certain non--conformal ${\cal N}=2$ theories, probe results were
presented in refs.\cite{bpp,ejp} which revealed new information about
the nature of the flow geometry and its singularities, and uncovered
new information about aspects of the gauge theory as captured by the
geometry.  As expected from the study of such non--conformal cases in
ref.\cite{jpp,cvj}, the appearance of an enhan\c con modifies
and clarifies a number of the purely supergravity conclusions.

There is a particular flow, found quite early in the
game\cite{freed1}, which represents a mass deformation of the theory
which flows in the IR to ${\cal N}=1$ supersymmetric $SU(N)$ gauge
theory ($N$ large) with two massless adjoint ``flavours''. It is a fixed
point theory, as known from the work of ref.\cite{robmatt} and
clarified in refs.\cite{lsflow,freed1}. There is a $U(1)$ R--symmetry
and an $SU(2)$ flavour symmetry.

In the language of the gravitational dual, the fact that the IR is
conformally invariant means that in both the $r=+\infty$ and the
$r=-\infty$ limits the geometry is asymptotically AdS$_5$, while the
transverse space changes from a round $S^5$ to a space with
$SU(2){\times}U(1)$ isometry.

Until relatively recently, the full ten dimensional geometry of this
solution was not known.  However, the full ten dimensional solution
representing this flow was presented in ref.\cite{pw2}, and we review
aspects of it in section~\ref{tendee}.

Since this deformation of the theory flows to a conformal field
theory, (we recall some of this in section~\ref{gaugetheory}), one
does not expect to have the sort of complicated probe behaviour which
was seen in the ${\cal N}=2$ non--conformal cases.  This expectation
is borne out by the explicit computations presented in this paper. We
probe the geometry with a D3--brane in section~\ref{probing} and find
the effective Lagrangian for the motion of the probe.  It is quite
well--behaved.  Specialising to the locus of points where the
potential vanishes, we see in section~\ref{coulomb} that the Coulomb
branch of moduli space is topologically ${\mathbf R}^4$, as it should
be, but deformed in a way which preserves only the $SU(2)_F\times
U(1)_R$ global symmetry. We compute and exhibit the metric on this
moduli space.

In fact, the probe computation yields a four dimensional moduli space
{\it everywhere along the flow}, including the far ultraviolet. This
is despite the fact that the supergravity solution asymptotes to
AdS$_5\times S^5$. This fits with the field theory: Since the probe at
radius $r$ computes the effective physics at some cutoff set by $u$ or
$r$, it knows about the true vacuum structure of the field theory,
which includes the fact that with the perturbation present, the moduli
space has four flat directions. In effect therefore, arbitrarily far
into the UV, the probe is sensitive to the IR physics, as we point out
precisely how the probe computation achieves this.

We amplify this point in the discussion of section~\ref{discuss}.
There, we point out the subtle fact that extracting the results of the
results of the probe computations, interpreting them in terms of a
four dimensional field theory, emphasises the characteristic
holographic nature of the D--brane probe.  While supergravity probes
see the features of local five or ten dimensional physics, the brane
probe physics has a four dimensional interpretation, and therefore is
sensitive to physics at very different radii, since these represent
different cutoff scales.  The results of the probe computation shows
just how this is achieved.  This is consistent with intuition about
the nature of effective Lagrangians, and is a key feature of how the
brane captures the holographic nature of the geometry/gauge theory
duality.
  
\section{The Ten Dimensional Solution}  
\label{tendee}  
The ten dimensional solutions computed in ref.\cite{pw2} describing
the gravity dual of ${\cal N}=4$ supersymmetric $SU(N)$ Yang--Mills
theory, mass deformed to ${\cal N}=1$ in the IR may be written as:
\begin{equation}  
ds^2_{10}=\Omega^2 ds^2_{1,4}+ds^2_{5}\ ,  
\labell{fullmetric}  
\end{equation}  
for the Einstein metric, where  
\begin{equation}  
ds^2_{1,4}=e^{2A(r)}\left(-dt^2+dx_1^2+dx_2^2+dx_3^2\right)+dr^2\ ,  
\labell{littlemetric}  
\end{equation}  
and\cite{cvetic,pw1}  
\begin{eqnarray}  
ds_5^2&=&{L^2}{\Omega^2\over\rho^2\cosh^2\chi}\left[{d\theta^2}
+\rho^6\cos^2\theta\left({\cosh\chi\over   
    {\bar X}_2}\sigma_3^2+{\sigma_1^2+\sigma_2^2\over 
{\bar X}_1}\right)\right.\nonumber\\  
&&\left.+{{\bar X}_2\cosh\chi\sin^2\theta\over {\bar X}^2_1} \left({d\phi}
+{\rho^{6}\sinh\chi \tanh\chi \cos^2\theta\over {\bar X}_2}
\sigma_3\right)^2\right]\ ,  
\labell{bigmetric}  
\end{eqnarray}  
with  
\begin{eqnarray}  
\Omega^2&=&{{\bar X}_1^{1/2}\cosh\chi\over \rho}\nonumber\\  
{\bar X}_1&=&\cos^2\theta+\rho^6\sin^2\theta\nonumber\\  
{\bar X}_2&=&{\rm sech}\chi\cos^2\theta+\rho^6\cosh\chi\sin^2\theta\ .  
\labell{warp}  
\end{eqnarray}  
The $\sigma_i$ are the standard $SU(2)$ left--invariant forms, the sum
of the squares of which give the standard metric on a round
three--sphere. They are normalised such that
$d\sigma_i=\epsilon_{ijk}\sigma_j\wedge\sigma_k$.  For future use, we
shall denote the coordinates on the $S^3$ as
$(\varphi_1,\varphi_2,\varphi_3)$.
  
The functions $\rho(r)\equiv e^{\alpha(r)}$ and $\chi(r)$ (to be discussed
more in detail shortly) which appear in the ten dimensional metric are
the supergravity scalars coupling to certain operators in the dual
gauge theory.  There is a one--parameter family of solutions for them
which gives therefore a family of supergravity solutions.
  
At $r\to\infty$, the UV, the various functions in the solution have
the following asymptotic values:
\begin{equation}  
\rho(r)\to 1\ ,\,\, \chi(r)\to 0\ ,\,\, A(r)\to {r\over L}\ .
\labell{asymptotes}  
\end{equation}  This gives AdS$_5\times S^5$, and the cosmological
constant is $\Lambda=-6/L^2$ where the normalisations are such that
the gauge theory and string theory quantities are related to them as:
\begin{eqnarray}
L=\alpha^{\prime1/2} (2g^2_{\rm
  YM}N)^{1/4}\ ;\qquad g^2_{\rm YM}=2\pi g_s\ .
\label{relations}
\end{eqnarray}
This limit defines
the $SO(6)$ symmetric critical point of the ${\cal N}=8$
supergravity scalar potential where all of the 42 scalars vanish.  At
the end of the flow, in the IR $r\to -\infty$, the functions asymptote to
the values:
\begin{equation}
\chi(r)\to{1\over2}\log 3\ ,
\,\,\, \alpha(r)\equiv\log\rho\to{1\over6}\ln 2\ ,\,\,\, A(r)\to
{2^{5/3}\over 3}{r\over L}\ .
\label{asympIR}
\end{equation}
which are the values defining another, $SU(2){\times}U(1)$ symmetric,
critical point of the scalar potential\cite{pilch}. It preserves only
${\cal N}=2$ supersymmetry of the maximal ${\cal N}=8$ for five
dimensional supergravity.
  
It is easily seen that the non--trivial radial dependences of
$\rho(r)$ and $\chi(r)$ deform the metric of the supergravity solution
from AdS$_5\times S^5$ at $r=+\infty$ where there is an obvious
$SO(6)$ symmetry (the round $S^5$ is restored), to a spacetime which
only has an $SU(2)\times U(1)$ symmetry, which is manifest in the
metric of equation~\reef{bigmetric}.\footnote{The $SU(2)$ is the
  left--invariance of the $\sigma_i$ and the $U(1)$ rotates $\sigma_1$
  into $\sigma_2$.  Actually the metric has an extra $U(1)$ symmetry,
  as $\frac{\partial}{\partial \phi}$ is also a Killing vector, but
  this is not a symmetry of the other fields in the full solution.}

The fields $\Phi$ and $C_{(0)}$, the ten dimensional dilaton and R--R
scalar, are gathered into a complex scalar field
$\lambda=C_{(0)}+ie^{-\Phi}$ on which $SL(2,{\mathbf Z})$ has a
natural action. This $SL(2,{\mathbf Z})$ is the duality symmetry of
the gauge theory in the UV (the dilaton is related to the gauge
coupling $g_{\rm YM}^{\phantom{2}}$, and the R--R scalar to the
$\Theta$--angle), and an action of it will be inherited by the gauge
theory in the IR. It was noticed in ref.\cite{pw2} that $\lambda$ is
constant all along the flow. This is the most obvious sign that the
geometry is going to be comparatively simple along the flow.

The non--zero parts of the two--form potential, $C_{(2)}$, and the
NS--NS two--form potential $B_{(2)}$ are also presented in
ref.\cite{pw2}, and they are non--trivial.  However, it turns out that
the probe brane we study is aligned in such a way that their
pull--back to the world--volume is exactly zero, so we do not exhibit
them here.

We will need, however, the explicit form for the R--R four form
potential $C_{(4)}$, to which the D3--brane naturally couples. The
derivatives of this field, which appear in the field strength, are
presented in ref.\cite{pw2}, and checks are made there on the mixed
second derivatives in order to ensure consistency. However, after some
algebra, it is possible to integrate the equations to yield a closed
form for the potential, the relevant part of which which we write
as\footnote{By ``relevant'', we mean the part which gets pulled back
  to the D--brane aligned parallel to the $(x^0,x^1,x^2,x^3)$
  directions.}:
\begin{eqnarray} 
C_{(4)}&=& -\frac{4}{g_s}w(r,\theta)\,
 dx_0\wedge dx_1\wedge dx_2 \wedge dx_3\ ,\nonumber \\
{\rm where}\quad  w(r,\theta)&=&
{e^{4A}\over8\rho^2}
[\rho^6\sin^2\theta(\cosh(2\chi)-3)-\cos^2\theta(1+\cosh(2\chi))]\ .  
\end{eqnarray}  
The radial dependences of the functions $\rho(r),\chi(r)$, and $A(r)$,
which appear in the ten dimensional solution, were found to be
governed by the reduction of the five dimensional supergravity
equations of motion to the following (recall that $\rho\equiv
e^\alpha$):
\begin{eqnarray}  
{d\rho\over dr}&=&{1\over 6L}
\left({\rho^6(\cosh(2\chi)-3)+2\cosh^2\chi\over\rho}\right)  
\nonumber \\  
{d\chi\over dr}&=&{1\over 2L}
\left({(\rho^6-2)\sinh(2\chi)\over \rho^2}\right)\nonumber \\  
{dA\over dr}&=&-{1\over 6L \rho^2}\left(\cosh(2\chi)(\rho^6-2)
-(3\rho^6+2)\right)\ . 
\labell{diffys}  
\end{eqnarray}

There is no known exact solution for these particular equations, but
much can be deduced about the structure of the solution by resorting
to numerical methods, as presented in ref.\cite{freed1}. It should be
noted that it is possible to extract the asymptotic UV ($r\to+\infty$)
behaviour of the fields $\chi(r)$ and $\alpha(r)=\log(\rho(r))$ is
given by:
\begin{equation}
\chi(r)\to a_0e^{-r/L}+\ldots\ ;
\qquad \alpha(r)\to {2\over3}a_0^2 {r\over L} e^{-2r/L}+{a_1\over\sqrt{6}}e^{-2r/L}
+\ldots
\label{asymptone}
\end{equation}
Crucially, the values of the constant\cite{pilch}
\begin{equation}
{\hat a}={a_1\over a_0^2}+\sqrt{8\over3}\log a_0
\end{equation}
characterise a family of different solutions for
$(\rho(r),\chi(r),A(r))$ representing different flows to the gauge
theory in the IR.  Meanwhile, in the IR ($r\to-\infty$) the asymptotic
behaviour is:
\begin{eqnarray}
&&\chi(r)\to {1\over2}\log 3-b_0e^{\lambda r/L}+\ldots\ ;
\qquad \alpha(r)\to {1\over6}\log 2-{\sqrt{7}-1\over6}b_0 e^{\lambda r/L}
+\ldots\ ,\nonumber \\
&&{\rm where}\quad \lambda={2^{5/3}\over3}(\sqrt{7}-1)\ .
\label{asympttwo}
\end{eqnarray}
At this end of the flow, there is also a combination which
characteristic of the flow, and this is $b_0a_0^\lambda$. This has
been pointed out in ref.\cite{freed1} as characterising the width of
the interpolating region.

The critical value\cite{freed1} ${\hat a}_c\simeq -1.4694$ represents the
particular flow which starts out at the ${\cal N}=4$ critical point
and ends precisely on the ${\cal N}=1$ critical point.  In
ref.\cite{gub2} it was proposed that the solutions with ${\hat
  a}>{\hat a}_c$ describe the gauge theory at different points on the
Coulomb branch of moduli space.  This fits with the fact that the
behaviour of $\chi$ is, according to the dictionary\cite{gkp,w1},
characteristic of an operator of dimension three representing mass
term (controlled by $a_0$), while that of $\alpha$ represents a {\it
  mixture} of both a dimension two mass operator (again through $a_0$)
and a vacuum expectation value (vev) of an operator of mass two
(through $a_1$). The combination ${\hat a}_c$ then, is pure mass and
no vev, while other values are a mixture of both. The vev is that of a
combination of massless fields which take us out onto the Coulomb
branch. We shall discuss this more in section~\ref{gaugetheory}.

For the flows with the ${\hat a}<{\hat a}_c$, the five dimensional
supergravity potential is no longer bounded above by the asymptotic UV
value and ref.~\cite{gub2} suggests that this makes them physically
unacceptable. They correspond to attempting to give a positive vev to
the massive field.

\section{The Gauge Theory}  
 \label{gaugetheory} 
The ${\cal N}=4$ supersymmetric Yang--Mills theory's gauge multiplet  
has bosonic fields $(A_\mu, X_i)$, $i=1,\ldots,6$, where the  
scalars $X_i$ transform as a vector of the $SO(6)$ R--symmetry, and  
fermions $\lambda_i$, $i=1,\ldots,4$ which transform as the $\bf 4$ of  
the $SU(4)$ covering group of $SO(6)$.  
  
In ${\cal N}=1$ language, there is a vector supermultiplet
$(A_\mu,\lambda_4)$, and three chiral multiplets made of a fermion and
a complex scalar ($k=1,2,3$): 
\begin{equation}
\Phi_k\equiv(\lambda_k,
\phi_k=X_{2k-1}+iX_{2k})\ ,
\end{equation}
and they have a superpotential
\begin{equation}
W=h{\rm Tr}(\Phi_3[\Phi_1,\Phi_2])\ ,
\end{equation}
where $h$ is related to $g_{\rm YM}$ in a specific way consistent with
superconformal symmetry.  We give a mass to $\Phi_3$,
\begin{equation}
L_{\rm ft}\to L_{\rm ft}+ \int d^2\theta \, {1\over2}m \Phi_3^2+{\rm h.c.}\ ,
\end{equation}
and then flow from the ${\cal N}=4$ gauge theory to the resulting
${\cal N}=1$ theory. This theory has ``matter'' multiplets in two
``flavours'', $\Phi_1$ and $\Phi_2$, transforming in the adjoint of
$SU(N)$.

The $SU(4)\simeq SO(6)$ R--symmetry of the ${\cal N}=4$ gauge theory is
broken to $SU(2)_F\times U(1)_R$, the latter being the R--symmetry of
the ${\cal N}=1$ theory, and the former a flavour symmetry under which
the matter multiplet forms a doublet.

So we switch on this small but relevant mass perturbation in the UV
(and possibly a vev of some of the massless fields too) and flow to
the IR. This maps to turning on certain scalar fields in the
supergravity, whose values asymptote to zero.  As one falls well below
the scale of the mass $m$ ---going to the IR--- these operators become
more relevant. In the supergravity solution, this corresponds to the
scalars being close to zero in the UV ($r\to+\infty$), developing a
non--trivial profile as a function of $r$, becoming more significantly
different from zero as one goes deeper into the IR, $r\to-\infty$.
This is precisely what is captured in equations~(\ref{asymptone})
and~(\ref{asympttwo}).  Finally, the supergravity equations of motion
require that there be a non--trivial back--reaction on the geometry,
which deforms the spacetime metric in a way given by $A(r)$, {\it
  etc.,} in section~\ref{tendee}.

Specifically, we must consider a combination of the
operators\cite{pilch}:
\begin{eqnarray}  
\alpha:\qquad&& \sum_{i=1}^4{\rm Tr} (X_iX_i)-2\sum_{i=5}^6{\rm Tr}  
(X_iX_i)  
\nonumber\\  
\chi:\qquad&& {\rm Tr} (\lambda_3\lambda_3+\phi_1[\phi_2,\phi_3])+{\rm h.c.},  
\end{eqnarray}  and we have listed the corresponding
scalar fields on the left. The specific nature of the terms is due to
operator mixing and the manner in which they combine to give the pure
mass deformation is discussed nicely in the review of ref.\cite{review}.

In fact, we can legitimately integrate out the massive scalar $\Phi_3$
at a low enough scale, and this results in the quartic superpotential
\begin{equation}
W={h^2\over 4m}{\rm Tr}([\Phi_1,\Phi_2]^2)\ ,
\end{equation}
which is in fact a marginal operator of the theory, defining a fixed
line of theories generated by varying its coefficient\cite{robmatt}.

The Coulomb branch moduli space of the ${\cal N}=1$ $SU(N)$ gauge
theory is parameterised by the vevs of the complex adjoint scalars
$\phi_{1,2}$ which set the potential ${\rm Tr}([\phi_1,\phi_2]^2)$ to
zero.  This generically breaks the theory to a product of $U(1)$'s.

We shall only probe a four dimensional subspace of the full moduli
space here since our moduli space is the space of allowed zero--cost
transverse movements of our single D3--brane probe. These directions
are parameterised by the scalars $(X_1,X_2,X_3,X_4)$, which make up
the complex doublet $(\phi_1,\phi_2)$.  Moving in that hyperplane
corresponds to the choice $\theta=0$ in the coordinates presented
earlier.

We shall find that the metric on this moduli space is very simple, and
is topologically ${\mathbf R}^4$. It will naturally inherit the
slightly squashed $S^3$ contained in the supergravity solution, and so
only have the action of an $SU(2)_F\times U(1)_R$ global symmetry of
the ${\cal N}=1$ gauge theory.

\section{Probing with a D3--brane}  
\label{probing}  
The uplifted geometry presented in ref.\cite{pw2} and listed in
section \ref{tendee} is given in the Einstein frame.  It is economical
to write the D3--brane world--volume action in terms of this metric:
\begin{eqnarray}  
S=-\tau_3\int_{{\cal M}_4} d^4\xi\,\,  
{\rm det}^{1/2}[{G}_{ab}+e^{-\Phi/2}{\cal  
  F}_{ab}]+\mu_3\int_{{\cal M}_4} \left(C_{(4)}+C_{(2)}\wedge{\cal{F}}
+{1\over2}C_{(0)}\, {\cal 
  F}\wedge {\cal F}\right), 
\label{actiontime} 
\end{eqnarray}  
where ${\cal F}_{ab}=B_{ab}+2\pi\alpha^\prime F_{ab}$, and ${\cal
  M}_4$ is the world--volume of the D3--brane, with coordinates
$\xi^0,\ldots,\xi^3$. As usual, the parameters $\mu_3$ and $\tau_3$
are the basic\cite{gojoe} R--R charge and tension of the D3--brane:
\begin{equation}
\mu_3 = \tau_3 g_s= (2\pi)^{-3}(\alpha^\prime)^{-2}\ .
\label{tau}
\end{equation}
Also, $G_{ab}$ and $B_{ab}$ are the pulls--back of the ten dimensional
metric (in Einstein frame) and the NS--NS two--form potential,
respectively which is defined as {\it e.g.}:
\begin{equation}  
G_{ab}=G_{\mu\nu}{\partial x^\mu \over\partial\xi^a}  
{\partial x^\nu \over\partial\xi^b}\ .  
\end{equation}  
  
Working in static gauge, we partition the spacetime coordinates,
$x^{\mu}$, as follows: $x^i=\{x^0,x^1,x^2,x^3\}$, and
$y^m=\{r,\theta,\phi,\varphi_1,\varphi_2,\varphi_3\}$. (The
$\varphi_i$ are angles on the deformed $S^3$ of section~\ref{tendee}.)
We choose static gauge as:
\begin{equation}  
x^0\equiv t=\xi^0\ ,\quad x^i=\xi^i\ ,\quad  y^m=y^m(t)\ .  
\end{equation}  
  
Putting everything together, we get the following result for the
effective Lagrangian for the probe moving {\it slowly} in the
transverse directions
$y^m=(r,\phi,\theta,\varphi_1,\varphi_2,\varphi_3)$ (we restrict
ourselves to considering $F_{ab}=0$ here):
\begin{eqnarray}  
{\cal L}\equiv T-V
={\tau_3\over2}
\Omega^2 e^{2A} G_{mn} {\dot y}^m{\dot y}^n
-\tau_3\sin^2\theta e^{4A}\rho^4(\cosh(2\chi)-1)\ .  
\label{theresult}
\end{eqnarray}  
Where the $G_{mn}$ refer to the Einstein frame metric components, and
we have neglected terms higher than quadratic order in the velocities
in constructing the kinetic term.

\section{Coulomb Branches}
\label{coulomb}
The logic of this whole approach is that the entire supergravity
solution is made of coincident D3--branes, carrying an $SU(N)$ gauge
theory. The probe computation represents the pulling of a test brane
out of the group, and exploring the background geometry and fields
produced by all of the others, which is the solution
(\ref{fullmetric}), with accompanying fields. This breaks $SU(N)\to
SU(N-1)\times U(1)$, and generically pulling them all apart would give
$U(1)^{N-1}$, although we focus on the result for one probe at a time.

Let us orient ourselves by recalling the UV case. We can obtain this
from our results by inserting the UV quantities given in equation
(\ref{asymptotes}) into equation (\ref{theresult}).  Our result {\it
  formally} (see below) gives the expected maximally supersymmetric
case of vanishing potential, giving flatness in all six transverse
directions to the brane\footnote{See for example, the review of this
  sort of computation in ref.\cite{primer} for a fuller discussion.}.
The metric on this moduli space is simply the flat metric on ${\mathbf
  R}^6$:
\begin{equation}
ds^2_{{\cal M}_{\rm UV}}={1\over 8\pi^2 g^2_{\rm YM}}[dv^2+v^2d\Omega_5^2]\ ,
\qquad{\rm with}\quad v={L\over\alpha^\prime}e^{{r/L}}\ ,
\label{round}
\end{equation}
where we have used the relations (\ref{relations}) and (\ref{tau}) and
defined the energy scale $v$. Here, $d\Omega^2_5$ is the metric
on a  round  $S^5$.

As stressed above, it is only formally correct to place those values into
the probe Lagrangian to get the answers above. We do {\it not} get
that result as the smooth endpoint of the flow. A more careful limit
of the potential should involve expanding the $\cosh(2\chi(r))$ term
for large $r$, inserting the behaviour given in (\ref{asymptone}).
This results in the behaviour
\begin{equation}
V(r)\sim e^{2r/L}\sin^2\theta\ ,
\label{veebig}
\end{equation}
showing that in fact only the directions specified by $\theta=0$ are
seen as flat directions in the UV limit, from the point of view of the
effective physics on the brane probe. So in fact, we should replace
$d\Omega_5^2$ in equation~(\ref{round}) by $d\Omega_3^2$, the metric
on a round $S^3$, since there are only four flat directions.  This
  is in contrast to what one would deduce locally from the full
  supergravity solution~(\ref{fullmetric}), which in the UV limit
goes smoothly to AdS$_5\times S^5$.

The point is that the Lagrangian on the probe at radius $r$ yields a
low energy effective action for the dual field theory below a cutoff
defined roughly by $v$ (or $r$). This effective action has knowledge
of the entire theory below that scale (at least\footnote{A {\it
    Wilsonian}\cite{wilson} exact effective action would of course
  also know about the UV physics. The precise relation of the
  effective action on a D--brane probe to a Wilsonian action deserves
  more careful consideration. See {\it e.g.}, refs.\cite{wilsonads}
  for work on effective actions (Wilsonian or otherwise) in the
  AdS/CFT Correspondence.  See also refs.\cite{tim} for the study of
  the Wilsonian exact effective action for gauge theory.}), especially
the far IR, and so the probe cannot lose sight of the fact that the
full theory is actually not the ${\cal N}=4$ gauge theory, but the
${\cal N}=1$ theory.  Note that this simple field theory fact has
somewhat profound holographic implications, and we shall emphasise
these points further in the discussion section.

So a general point on the flow has $\theta=0$ as the family of flat
directions\footnote{The case $\rho=0$, which is $\alpha=-\infty$, lies
  outside the physically allowed values of the flow.}.  This moduli
space is the Coulomb branch of the gauge theory anywhere along the
flow. We see that we have movement on the (squashed) $S^3$, with
coordinates $(\varphi_1,\varphi_2,\varphi_3)$, and the radial
direction~$r$. These give an~${\mathbf R}^4$, topologically, which is
appropriate to the fact that we have two complex scalar fields in the
adjoint, $\phi_1$ and $\phi_2$, whose vevs we can explore.  The metric
on this moduli space for arbitrary $(r,\varphi_1,\varphi_2,\varphi_3)$
is:
\begin{equation}
ds^2={\tau_3\over2}{\cosh^2\chi\over\rho^2}e^{2A}
dr^2+{\tau_3\over2}{L^2}{e^{2A}}\rho^2
\left({\cosh^2\!\chi}\,\sigma_3^2+\sigma_1^2+\sigma_2^2\right)\ , 
\end{equation} 
whose behaviour can be verified by a numerical study of the
flow equations for the functions $(\chi(r),\rho(r), A(r))$.  

We can study this metric in the IR limit $r\to-\infty$. Inserting the
IR values of the functions (see equation (\ref{asympIR})), using the
relations in equations (\ref{relations}) and (\ref{tau}), and
defining:
\begin{eqnarray}
u={\rho_0L\over\alpha^\prime}e^{{r/ \ell}}\ ,\qquad \ell={3\over 2^{5/3}}
 L\ ,\qquad
\qquad\rho_0\equiv\rho_{\rm
  IR}=2^{1/6}
\end{eqnarray}
we get this extremely pleasing form for the metric:
\begin{equation}
ds^2_{{\cal M}_{\rm IR}}={1\over 8\pi^2 g_{\rm YM}^2}\left[{3\over4} du^2+
 u^2\left({4\over 3}\sigma_3^2+
{\sigma_1^2+\sigma_2^2}\right)\right]\ .
\label{squashed}
\end{equation}
There is a conical singularity at the origin, as a result of the 4/3
coefficient of $\sigma_3$.  The $S^3$'s in constant~$r$ or~$u$ radial
slices are squashed by the presence of $\cosh^2\chi$ (which is equal
to 4/3 in the IR) instead of unity.  Such a deformation is to be
expected, since the residual isometry is $SU(2)_F\times U(1)_R$, the
global symmetry of the gauge theory.  It would be interesting to find
the gauge theory interpretation of this metric, including the
singularity. The radial coordinate~$u$ that we have chosen in the IR
is motivated by the choice (\ref{round}) of radial coordinate $v$ in
the~UV, where, because of the extra supersymmetry, the resulting $v$
dependence corresponds to the vanishing of the $\beta$--function.  It
is interesting that we have a similar dependence here since under
$u\to\lambda u$, the metric (\ref{squashed}) recales by $\lambda^2$,
for $\lambda$ real.  This is suggestive of conformal invariance, however
we cannot directly conclude anything about the $\beta$--function since
we only have ${\cal N}=1$ supersymmetry, which provides no relation
between the kinetic term and the gauge coupling.

Further comparison to the gauge theory (yielding understanding of {\it
  e.g.,} the conical singularity) perhaps requires finding coordinates
which are better adapted to the fixed point theory. This is the
subject of ongoing research.

\section{Discussion and Holographic Reflections}
\label{discuss}
In summary, we have studied the results of probing a particular ten
dimensional type~IIB supergravity solution with a D3--brane. The
solution has the dual interpretation as a renormalisation group flow
from the maximally supersymmetric conformally invariant ${\cal N}{=}4$
$SU(N)$ gauge theory to a conformally invariant ${\cal N}{=}1$
supersymmetric $SU(N)$ gauge theory with two massless matter flavours
in the adjoint.  Both gauge theories are of course at strong 't Hooft
coupling, $g_{\rm YM}^2N$, with large~$N$ and small~$g_{\rm YM}$.

This sort of flow is expected to be simple, since the beginning and
ends of the trajectory are conformal, and the spacetime is AdS (times
a compact manifold) at both ends. This is reflected in the good
behaviour of the probe problem everywhere along the flow, in
necessary contrast to the results obtained for non--conformal flows
in refs.\cite{bpp,ejp}.

The piece of the Coulomb branch that the computation yields is
consistent with expected properties of the gauge theory in the IR:
there is a simple scaling property of the metric, with an isometry
corresponding to the $SU(2)_F\times U(1)_R$ global symmetry. One can
also solve for these flat directions anywhere along the flow, which is
interesting.

Before closing we would like to highlight a remarkable holographic
feature revealed by the probe computation. Recall that the
supergravity goes back to the complete AdS$_5\times S^5$ solution in
the UV ($r\to+\infty$) limit, and forgets about the perturbation by
the relevant operators to any desired accuracy if one goes to large
enough $r$.  This is not the case for the probe result. The effective
Lagrangian of the probe does not forget about the perturbation, since,
as we pointed out in the previous section, an $\exp(2r/L)$ factor
prevents us matching smoothly onto the result for a probe of pure AdS.

This is crucial and we emphasise it: {\it The ten dimensional geometry
  of the flow solution becomes arbitrarily close to that of pure
  AdS$_5\times S^5$ in the UV, but the physics of the D3--Brane
  probing the flow geometry does not approach that seen by a D3--brane
  in pure AdS$_5\times S^5$.}

This is counterintuitive from the point of view of the physics of
local probes: One would expect that locally, for large positive $r$, a
probe (such as one which knows only about the local metric) cannot
tell that it is not in simple AdS$_5\times S^5$, and the small
deviations do not matter. This is not so for a D3--brane probe: It can
tell the difference arbitrarily far into the UV ($r\to+\infty$),
because of the $\exp(2r/L)$ factor discussed around
equation~(\ref{veebig}).\footnote{See {\it e.g.}  refs.\cite{iruv} for
  some discussion of the issue of probes in AdS and holography.}

{}From the point of view of the four dimensional gauge theory physics,
however, this is precisely the right expectation.
The effective physics computed for the probe at any radius is an
expression of the physics of the gauge theory at the effective scale
defined by $r$ (or $u$). So the probe's effective Lagrangian should
capture the deep infrared physics anywhere along the flow, and should
in particular know that it is  the ${\cal N}=1$ theory and not the
${\cal N}=4$. Therefore, no matter how far one runs into the UV (large
$r$), the probe should be sensitive to the physics present in the IR
(small $r$). A supergravity probe cannot manage this, but a D3--brane
can, as shown in the simple, clear example of this paper.

Running the logic the other way, if one had two supergravity
geometries which were similar in the IR but quite different in the UV,
one would expect that the probe physics would be less sensitive to the
differences in the UV, since those are short distance details that do
not crucially affect the low energy effective action.

We remark that it is certainly worth studying these probe techniques
further, applying them to other new (inevitably more complicated) ten
dimensional geometries which are appearing on the market regularly:
they offer complementary insights into the physics that gauge/geometry
dualities can teach us, and are a useful route by which we can bridge
the gap between purely supergravity techniques and the full
superstring technology needed to fully investigate these backgrounds.

\section*{Acknowledgements} 
We thank Nick Evans and Michela Petrini for comments, and Douglas
Smith for discussions.  DCP and KJL would like to thank the Institut
Henri Poincar\'e, Paris for hospitality during the completion of this
work.  They would also like to thank Ian Davies, James Gregory and
Antonia Padilla for helpful discussions.  This paper is report number
DTP/00/97 at the CPT, Durham.
 


\end{document}